\let\includefigures=\iftrue
\let\useblackboard=\iftrue
\newfam\black
\input harvmac

\noblackbox

\includefigures
\message{If you do not have epsf.tex (to include figures),}
\message{change the option at the top of the tex file.}
\input epsf
\def\figin{\epsfcheck\figin}\def\figins{\epsfcheck\figins}
\def\epsfcheck{\ifx\epsfbox\UnDeFiNeD
\message{(NO epsf.tex, FIGURES WILL BE IGNORED)}
\gdef\figin##1{\vskip2in}\gdef\figins##1{\hskip.5in}
\else\message{(FIGURES WILL BE INCLUDED)}%
\gdef\figin##1{##1}\gdef\figins##1{##1}\fi}
\def\DefWarn#1{}
\def\figinsert{\goodbreak\midinsert}
\def\ifig#1#2#3{\DefWarn#1\xdef#1{fig.~\the\figno}
\writedef{#1\leftbracket fig.\noexpand~\the\figno}%
\figinsert\figin{\centerline{#3}}\medskip\centerline{\vbox{
\baselineskip12pt\advance\hsize by -1truein
\noindent\footnotefont{\bf Fig.~\the\figno:} #2}}
\bigskip\endinsert\global\advance\figno by1}
\else
\def\ifig#1#2#3{\xdef#1{fig.~\the\figno}
\writedef{#1\leftbracket fig.\noexpand~\the\figno}%
\global\advance\figno by1}
\fi
%

\useblackboard
\message{If you do not have msbm (blackboard bold) fonts,}
\message{change the option at the top of the tex file.}
\font\blackboard=msbm10 scaled \magstep1
\font\blackboards=msbm7
\font\blackboardss=msbm5
\textfont\black=\blackboard
\scriptfont\black=\blackboards
\scriptscriptfont\black=\blackboardss

\else

\fi
%
\def\yboxit#1#2{\vbox{\hrule height #1 \hbox{\vrule width #1
\vbox{#2}\vrule width #1 }\hrule height #1 }}
\def\fillbox#1{\hbox to #1{\vbox to #1{\vfil}\hfil}}
\def\ybox{{\lower 1.3pt \yboxit{0.4pt}{\fillbox{8pt}}\hskip-0.2pt}}
%
%

\def\xit{\xi}


\def\comments#1{}

\def\CA{{\cal A}}

\def\CN{{\cal N}}
\def\CO{{\cal O}}

\def\CL{{\cal L}}


\def\II{\relax{I\kern-.10em I}}

\def\IZ{\relax\ifmmode\mathchoice
{\hbox{\cmss Z\kern-.4em Z}}{\hbox{\cmss Z\kern-.4em Z}}
{\lower.9pt\hbox{\cmsss Z\kern-.4em Z}}
{\lower1.2pt\hbox{\cmsss Z\kern-.4em Z}}
\else{\cmss Z\kern-.4emZ}\fi}
\def\IB{\relax{\rm I\kern-.18em B}}
\def\IC{{\relax\hbox{$\inbar\kern-.3em{\rm C}$}}}
\def\ID{\relax{\rm I\kern-.18em D}}
\def\IE{\relax{\rm I\kern-.18em E}}
\def\IF{\relax{\rm I\kern-.18em F}}
\def\IG{\relax\hbox{$\inbar\kern-.3em{\rm G}$}}
\def\IGa{\relax\hbox{${\rm I}\kern-.18em\Gamma$}}
\def\IH{\relax{\rm I\kern-.18em H}}
\def\II{\relax{\rm I\kern-.18em I}}
\def\IK{\relax{\rm I\kern-.18em K}}
\def\IP{\relax{\rm I\kern-.18em P}}

%

\def\inbar{\,\vrule height1.5ex width.4pt depth0pt}

\font\cmss=cmss10 
\def\IR{\relax{\rm I\kern-.18em R}}

\def\dpr{^{\prime\dagger}}
%


%

\def\lp10{\ell_p^{10}}
\def\lp11{\ell_p^{11}}
\def\R11{R_{11}}

\def\frac#1#2{{#1 \over #2}}

\def\dS{\partial \Sigma}


\def\uu{^}
\def\ll{_}

\def\b{\beta}
\def\s{\sigma}

\def\pr{^\prime}

\def\th{\theta}

\def\dag{^\dagger}

\def\d{\delta}

\def\sqd{^2}

\def\hsk{\hskip .5in}


\def\1dag{^{1\dagger}}
\def\2dag{^{2\dagger}}

\def\R#1#2#3{{{R_{#1}}^{#2}}_{#3}}

\def\lsq{\left [}
\def\rsq{\right ]}

\def\Pt{{\wp}}
\def\bpr{\beta \dpr}
\def\bprd{\beta \pr}

\def\ie{{\it i.e.}}
\def\eg{{\it e.g.}}
\def\cf{{\it c.f.}}

\hyphenation{Di-men-sion-al}

\def\np{{\it Nucl. Phys.}}
\def\prl{{\it Phys. Rev. Lett.}}
\def\prev{{\it Phys. Rev.}}

\def\mpl{{\it Mod. Phys. Lett.}}

\def\jhep{{\it J. High Energy Phys.}}
\def\ijmp{{\it Int. J. Mod. Phys.}}

\lref\lvw{W. Lerche, C. Vafa and N.P. Warner,
``Chiral rings in $\CN=2$ superconformal theories,''
\np\ {\bf B324}\ (1989) 427.}
\lref\wityang{E. Witten, ``On the Landau-Ginzburg description of
N=2 minimal models,'' \ijmp\ {\bf A9} (1994) 4783;
hep-th/9304026.}
\lref\vafa{C. Vafa, ``String vacua and orbifoldized LG models'',
\mpl\ {\bf A4} (1989) 1169\semi
K. Intriligator and C. Vafa, ``Landau-Ginzburg orbifolds'', \np\
{\bf B339} (1990) 95.}

\lref\warner{N. Warner, ``Supersymmetry in boundary integrable models,''
\np\ {\bf B450} (1995) 663; hep-th/9506064}

\lref\phases{E. Witten, ``Phases of N=2 theories in two dimensions,''
\np\ {\bf B403} (1993) 159; hep-th/9301042.}
\lref\agm{P. Aspinwall, B. Greene and D. Morrison,
``Calabi-Yau moduli space, mirror manifolds and space-time topology
change in string theory,'' \np\ {\bf B416} (1994) 414; hep-th/9309097. }
\lref\cylg{B. Greene, C. Vafa and N. Warner,
``Calabi-Yau manifolds and renormalization group flows,''
\np\ {\bf B324} (1989) 371\semi
E. Martinec, ``Criticality, Catastrophes
and Compactifications,'' in Brink, L. (ed.) et al., {\it Physics
and Mathematics of Strings} 383. }
\lref\web{T.-M. Chiang and B.R. Greene, ``Phases of
mirror symmetry,'' Strings '95 proceedings
{\it Future Perspectives in String Theory} (World Scientific, New Jersey)
1996; hep-th/9509049.}

\lref\Gepner{D. Gepner, ``Exactly Solvable String Compactifications
on Manifolds of $SU(N)$ Holonomy,'' Phys. Lett. {\bf B199} (1987) 380.}

\lref\hklm{S. Kachru, S. Hellerman, A. Lawrence and J. McGreevy, to
appear, as presented
by S. Kachru at Strings 2000, Michigan.}
\lref\horivafa{K. Hori, A. Iqbal and C. Vafa,
``D-Branes and mirror symmetry''; hep-th/0005247.}
\lref\diacdoug{D.-E. Diaconescu and M.R. Douglas, ``D-branes
on stringy Calabi-Yau manifolds''; hep-th/0006224.}
\lref\govinda{S. Govindarajan, T. Jayaraman, and T. Sarkar,
``On D-branes from gauged linear sigma models''; hep-th/0007075.}
\lref\mayr{P. Mayr, ``Phases of
supersymmetric D-branes on Kaehler manifolds and the McKay correspondence,''
hep-th/0010223.}
\lref\hori{K. Hori, ``Linear models of supersymmetric D-branes,''
hep-th/0012179.}

\lref\shamitjohn{S. Kachru and J. McGreevy, ``Supersymmetric
three-cycles and supersymmetry breaking,'' \prev\ {\bf D61} (2000)
026001; hep-th/9908135.}
\lref\kklmone{S. Kachru, S. Katz, A. Lawrence and J. McGreevy, ``Open
String Instantons and Superpotentials,'' \prev\
{\bf D62} (2000) 026001;  hep-th/9912151.}
\lref\kklmtwo{S. Kachru, S. Katz, A. Lawrence and J. McGreevy, ``Mirror
symmetry for open strings''; hep-th/0006047.}

\lref\bbs{K.~Becker, M.~Becker and A.~Strominger,
``Five-branes, membranes and nonperturbative string theory,"
\np\ {\bf B456} (1995) 130; hep-th/9507158.}
\lref\ooy{H. Ooguri, Y. Oz and Z. Yin, ``D-branes
on Calabi-Yau spaces and their mirrors,''
\np\ {\bf B477} (1996) 407; hep-th/9606112.}
\lref\bpsalg{J.A. Harvey and G. Moore,
``On the algebras of BPS states,'' Comm. Math. Phys. {\bf 197}
(1998) 489, hep-th/9609017.}
\lref\bdlr{I. Brunner, M.R. Douglas, A. Lawrence and
C. R\"omelsberger, ``D-branes on the Quintic,''
\jhep\ {\bf 0008} (2000) 015; hep-th/9906200.}
\lref\morrpless{D. Morrison and M.R. Plesser, ``Summing the
Instantons: Quantum Cohomology and Mirror Symmetry in Toric Varieties,''
Nucl. Phys. {\bf B440} (1995) 279, hep-th/9412236.}
\lref\recentmrd{M. Douglas, ``D-branes, Categories, and $\CN=1$
Supersymmetry,''
hep-th/0011017.}

\lref\muto{T. Muto, ``D-branes at orbifolds and
topology change,'' \np\ {\bf 521}\ (1998) 183;
hep-th/9711090.}
\lref\dflop{B.R. Greene, ``D-brane topology changing transitions,''
\np\ {\bf B525}\ (1998) 284; hep-th/9711124.}
\lref\mrdstab{M.R. Douglas, B. Fiol and C. R\"omelsberger,
``Stability and BPS branes''; hep-th/0002037.}
\lref\mrdspectra{M.R. Douglas, B. Fiol and C. R\"omelsberger,
``The spectrum of BPS branes on a noncompact Calabi-Yau'';
hep-th/0003263.}

\lref\dk{J. Distler and S. Kachru, ``(0,2) Landau-Ginzburg Theory,''
{\it Nucl. Phys.} {\bf B413} (1993) 213, hep-th/9309110.}
\lref\trieste{J. Distler, ``Notes on (0,2) superconformal
field theories,'' Trieste HEP Cosmology 1994:0322-351, hep-th/9502012.}
\lref\kw{S. Kachru and E. Witten, ``Computing the Complete Massless
Spectrum of a Landau-Ginzburg Orbifold,'' hep-th/9307038.}
\lref\evaed{E. Silverstein and E. Witten, ``Criteria
for conformal invariance of $(0,2)$ models,''
\np\ {\bf 444} (1995) 161; hep-th/9503212.}
\lref\resolving{J. Distler, B. Greene, D. Morrison, ``Resolving
singularities in (0,2) models,'' hep-th/9605222.}

\lref\peskin{E. Mirabelli and M. Peskin, ``Transmission of Supersymmetry
Breaking from a 4-Dimensional Boundary,'' Phys. Rev. {\bf D58} (1998)
065002, hep-th/9712214.}
\lref\matrixcy{S. Kachru, A. Lawrence and E. Silverstein,
``On the matrix description of Calabi-Yau compactifications,''
\prl\ {\bf 80}\ (1998) 2996; hep-th/9712223.}
\lref\wittmirr{E. Witten, ``Mirror manifolds and
topological field theory,'' in {\it Mirror Symmetry I},
S.-T. Yau (ed.), American Mathematical Society (1998),
hep-th/9112056.}
\lref\boundarysg{P. Fendley, H. Saleur and
N.P. Warner, ``Exact solution of a massless
scalar field with a relevant boundary interaction,''
\np\ {\bf B430}\ (1994) 577; hep-th/9406125.}
\lref\gandh{P. Griffiths and J. Harris, ``Principles of Algebraic Geometry,''
{\it John Wiley and Sons, Inc.} (1978).}
\lref\hartshorne{R. Hartshorne, ``Algebraic Geometry,''
{\it Springer} (1977).}
\lref\sft{
{\it e.g. } E. Witten, ``On background independent open-string field theory,''
\prev\ {\bf D46}(1992) 5467, hep-th/9208027;
S. Shatashvili, ``Comment on the background independent open
string theory,'' {\it Phys. Lett.} {\bf B311} (1993) 83, hep-th/9303143;
S. Shatashvili, ``On the problems with background independence in
string theory,'' {\it Algebra and Anal. v. 6} (1994) 215, hep-th/9311177;
A. Gerasimov, S. Shatashvili, ``On exact tachyon potential in open
string field theory,'' JHEP {\bf 0010} (2000) 034;
J. Harvey, D. Kutasov and E. Martinec, ``On the relevance of tachyons,''
hep-th/0003101;
D. Kutasov, M. Marino and G. Moore,
``Some exact results on tachyon condensation in
string field theory,''
JHEP{\bf 0010}(2000) 045,
hep-th/0009148.}
\lref\biglenny{D. Bigatti, L. Susskind,
``Magnetic fields, branes and noncommutative geometry,''
\prev\ {\bf D62} (2000) 066004, hep-th/9908056.}
\lref\AspinwallHE{
P.~S.~Aspinwall and R.~Y.~Donagi,
``The heterotic string, the tangent bundle, and derived categories,''
Adv.\ Theor.\ Math.\ Phys.\  {\bf 2}, 1041 (1998);
hep-th/9806094.
}
\lref\SharpeQZ{
E.~R.~Sharpe,
``D-branes, derived categories, and Grothendieck groups,''
Nucl.\ Phys.\ B {\bf 561}, 433 (1999); hep-th/9902116.
}
\lref\kontsevich{
M. Kontsevich, ``Homological Algebra of Mirror Symmetry,''
alg-geom/9411018.}

\Title{\vbox{\baselineskip12pt\hbox{hep-th/0104100}
\hbox{SLAC-PUB-8808}\hbox{SU-ITP-01/19}}}
{\vbox{
\centerline{Linear sigma model toolshed
}
\bigskip
\centerline{for D-brane physics}}}
\bigskip
\bigskip
\centerline{Simeon Hellerman$^{1,2}$ and John McGreevy$^{1}$}
\bigskip
\centerline{$^{1}${\it Department of Physics, Stanford University,
Stanford, CA 94305}}
\smallskip
\centerline{$^{2}${\it SLAC Theory Group, MS 81, PO Box 4349,
Stanford, CA 94309}}
\bigskip
\bigskip
\noindent
Building on earlier work \refs{\hklm},
we construct linear sigma models for strings on curved spaces
in the presence of branes.
Our models include  an
extremely general class of brane-worldvolume gauge field configurations.
We explain in an accessible manner the mathematical
ideas which suggest appropriate worldsheet interactions for generating a
given open string background.  This construction provides an explanation
for the appearance of the derived category in D-brane physics
complementary to that of \recentmrd.

\bigskip
\Date{March 2001}

\newsec{Introduction}

\it
\noindent
Main Entry: de $\cdot$ noue $\cdot$ ment

\noindent
Function: noun
\noindent
\item{1.}  the final outcome of the main dramatic complication in a literary work
\item{2.}  the resolution of a complex sequence of events
\rm
\bigskip

The study of supersymmetric D-branes in curved spaces is a
dual-purpose endeavor.  On one hand, these objects provide new probes of the
stringy physics which was uncovered
principally by the application of mirror symmetry.
On the other hand, when coupled with orientifolds, they provide a
large, relatively uncharted class of
quasi-realistic string vacua.

Thus far, only a small part of the spectrum
of branes on Calabi-Yau (CY) manifolds
has been studied.  While a classification
of these objects, even in the geometrical limit, is lacking,
an avenue toward systematic exploration is provided by
the fact that all bundles have a description
in terms of simpler components.  Such
a description, called a ``resolution''
(see \gandh\ or \hartshorne\ Chapter II, Corollary 5.18)
is a sequence of maps between
sums of rank-one bundles
which encodes the non-triviality of the bundle of interest.
As a means of clearly laying out the space of D-brane states,
these sequences appear promising \recentmrd.
It should be possible
to use this sequence data
to make a useful model of the stringy physics.

A tractable description of the
conformal field theory (CFT) describing
the string dynamics on a CY is lacking
(for any branes, or even in their absence)
except for special values of the closed string moduli.
A useful model of much of the physics at all points
in moduli space is provided by linear sigma models (LSM's)\phases.
Such models are two-dimensional quantum field theories
which approach the desired CFT at long distances.
Linear sigma models are in many cases the only available tools
for extracting information about string
backgrounds in the small-volume region, including
the persistence of such backgrounds beyond perturbation theory.
The LSM is the ideal framework to

\smallskip
{\centerline {\it  HARNESS THE AWESOME POWER OF HOMOLOGICAL ALGEBRA.}}
\smallskip

\noindent
Already the shortest such
sequences (``monads'') have appeared in linear sigma model
constructions, first for heterotic strings \refs{\phases, \dk}, and
more recently for open strings \hklm.

In this paper, we make linear models for bundles whose
resolution has an arbitrary number of nodes\foot{It
was suggested recently \recentmrd\ that this would be
useful.}.  More generally, the models we make
generate any bundle which can be realized as
the cohomology of a sequence of sums of line
bundles.

The sequences which encode the data for these models are
the same ones that appear as building-blocks for the derived
category of \recentmrd.\foot{
Early work in this direction includes \refs{\kontsevich, \AspinwallHE, \SharpeQZ}.
}
As in other incarnations of open string field theory
\sft,
the condensation of spacetime tachyons is manifested
on the string worldsheet
as flow to the infrared.
In our construction, the equivalence of quasi-isomorphic
complexes is a consequence of universality
in the sense of the renormalization group.

There a second way
in which our work relates to \recentmrd\
which we explain in the concluding section.

In addition to any motivation from
recent exciting ideas about abstract descriptions of D-brane spectra,
our construction would be needed
for an attempt to make contact
with phenomenology through this type
of four-dimensional $\CN = 1$ string vacuum.
Certainly any systematic exploration of this
class of vacua would incorporate the more general
models we construct.

The plan of the paper is as follows.
In \S2 we (p)review the construction \hklm\ of a linear sigma model
coupling open strings to B-type branes, and
more specifically to a bundle defined by a one-step sequence.
In \S3 we explain why this construction fails for more general bundles
whose resolutions do not terminate after one step, and
explain our solution.
In \S4 we discuss a linear model for an even broader class of bundles,
namely those which are not pullbacks of bundles on the
ambient space in which the CY is embedded.
In \S5 we explore the relation of our linear models to possible
linear models for heterotic strings whose left-moving fermions couple
to these more general bundles.
We close with a discussion of
applications and the extension of
this set of ideas to CY's and branes which are not complete
intersections.
In appendix A we give more details about our models.  In
appendix B we describe another formulation of our models,
of which the one in the body of the paper is a gauge-fixed version.

Other work on linear models for open strings includes
\refs{\horivafa, \diacdoug, \govinda, \mayr, \hori}.

\newsec{Brief review of the construction for one step}

We present this discussion in the context of
a CY hypersurface in projective space, but the generalization
to any complete intersection in a toric variety
should be clear.
The simplest example of the
monad construction defines a holomorphic vector bundle, $V$,
from the sequence of maps
$$ 0 \to V \to E_1 \equiv \bigoplus_{a_1} \CO(n_{a_1})
{\buildrel d_1 \over\longrightarrow} E_2 \equiv \bigoplus_{a_2}
\CO(n_{a_2}) \to 0.$$
The first map is inclusion and the second
map is
$$
\eqalign{  d_1: & E_1 \to E_2 \cr
    & \beta^{a_1} \mapsto \sum d^{ a_2}_{1 ~ a_1}(\phi) \beta^{a_1}.
}
$$
This sequence is {\it exact} in the sense that
the kernel of one map is the image of the previous one.
For our purposes, the bundles $E_l$ are sums of powers of
the hyperplane bundle over a projective space.  We want to
consider type II string theory on a Calabi-Yau hypersurface, $X$,
in the projective space.
A linear sigma model which realizes open strings ending
on a brane wrapping $X$ with gauge bundle $V$
works as follows \hklm.

The bulk fields are the same as in the (2,2) linear model for $X$.
For ease of exposition we present the model
for a hypersurface in $\IP^4$ defined by
a homogeneous polynomial $G$ of degree $d=5$.
Then one has a single $U(1)$ gauge multiplet
under which the chiral multiplets $\Phi^i,~i=1\dots 5$
carry charge $1$ and $P$ carries charge $-5$.
There is a (2,2) superpotential $W_{bulk} = P G(\Phi)$.
In accordance with the notation of the relevant (B-type)
topological field theory, we refer to the fermions
in a bulk chiral multiplet as
$$ \Theta = {1 \over \sqrt{2}} (\psi_+ - \psi_-), ~~~ \eta =
{1 \over \sqrt{2}} (\psi_+ + \psi_-).$$
We refer to \phases\ for more details.

We give these fields couplings to
boundary matter
which respect B-type supersymmetry.
By B-type supersymmetry we mean
a subalgebra of the bulk (2,2) supersymmetry
generated by a linear combination
of $Q_+$ and $Q_-$ rather than
a linear combination of $Q_+$ and $Q^\dagger_-$.
Boundary conditions preserving such a subalgebra
are associated with branes
of even codimension in the CY \ooy.
We call the conserved
supercharges $Q$ and $Q^\dagger$.
The representation theory of their
algebra is similar to that of (0,2) two-dimensional supersymmetry and
is worked out in \hklm.
The two kinds of multiplets we will need are:
Fermi multiplets, $\beta$, have a fermion as their lowest component, and
chiral multiplets, $\wp$, have a boson as their lowest component.
Both satisfy either the chiral constraint, $Q^\dagger (\beta, \wp) =0 $,
or some deformation thereof.  The consistency conditions
on such deformations will play a key role in building the more
general bundles.

\subsec{About kinetic terms}

A Fermi multiplet contains a complex fermion, $\beta$, and an auxiliary
complex boson, $b$.  A supersymmetric kinetic term is \foot{We use
gauge-covariantized supercharges, so we
omit the usual $e^{qV}$'s.}

$$
\int d\sqd \th \b\dag\b = b\dag b  - i (\nabla\ll 0 \b\dag)
\b
+ i \b\dag (\nabla\ll 0 \b )
$$
where
$$ \nabla_0 \equiv {i \over 2} \{ Q , Q\dag \} .$$

A boundary chiral multiplet contains a complex boson, $\wp$,
and a fermion $\xi$.  In addition to a usual kinetic
term of the form

$$ \int d^2 \theta \wp^\dagger \nabla_0 \wp ,$$
we add a magnetic field term:

$$\int d^2 \theta B \wp^\dagger \wp =
        B \xi^\dagger \xi + B \wp^\dagger \nabla_0 \wp  - (\nabla_0 B
\wp^\dagger) \wp$$
where $B$ is a constant.
Consider a limit where $B$ is big
so that we can ignore the kinetic terms for $\wp$, \cf\ \eg\
\biglenny.  This
approximation can be justified by the fact
that this term is less irrelevant than the kinetic one.
In that case, the momentum conjugate to $\wp$ is $i B \wp^\dagger$.
So canonical quantization gives
$$ [ \wp, \wp^\dagger ] = 1/B .$$
The coupling to the magnetic field  masses up the fermion $\xi$ and
it halves the number of $\wp$ degrees of freedom - i.e. the set of
states
made by $\wp$ is now just the Hilbert space of a harmonic oscillator.
It makes the $\wp$ multiplet into an exact bosonic analog of a Fermi
multiplet.  Since we work in an approximation where the magnetic field
term
dominates the kinetic term, we will drop the usual kinetic term altogether and
adopt
a normalization for $\wp$ such that $B\equiv 1$.

\subsec{The model for a monad}

On the boundary of the string, introduce
Fermi multiplets, $\beta^{a_1}$, which live in the bundle $E_1$, and
chiral multiplets, $\wp_{a_2}$, which
live in the bundle $E_2^\star$.
In this case the half-superspace integrand which generates the bundle $V$
is
$$ W = \wp_{a_2} d_{1~a_1}^{a_2} \beta^{a_1}$$
which we abbreviate as
$$ W = \wp d\ll 1 \beta$$
(We will suppress indices whenever possible!)

After integrating out the auxiliary bosons in $\beta_a$, this leads to
the following interactions on the boundary
$$ \CL = \dots + \xi_{a_2} d_{1~a_1}^{a_2} \beta^{a_1} + h.c.
+ \sum_{a_2} |\wp_{a_2} d_{1~a_1}^{a_2}|^2 $$

The finishing touch we need to put on the monad model is to
implement a charge projection on the boundary
which guarantees the right number of Chan-Paton (CP)
states -- i.e., that only one Chan-Paton fermion at a time will be excited, no
more and no less.  Specifically, we
gauge the symmetry which acts only on boundary fields, as
$$
\eqalign{
\wp &\mapsto e^{i \gamma} \wp \cr
\beta &\mapsto e^{-i \gamma} \beta.
}
$$
This is done by adding a supersymmetry singlet,
one-dimensional gauge field $a_0$ on the boundary,
which couples as
$$ \CL_a = \int_{\dS} a_0 (j_S - 1) $$
where $j_S$ is the boundary symmetry charge.  Since each excitation of
the $\beta$ fermions is in effect a 'quark' at the left endpoint of the
open string (or an 'antiquark' at the right endpoint), the charge
projection
restricts the system to a subsector of the Hilbert space where the open
string
worldsheet couples to the gauge field as a section of $V\otimes V^*$
\refs{\hklm}.

\newsec{Any resolution}

Consider a bundle $V$ over a Calabi-Yau manifold which
is a hypersurface in projective space.
Such a bundle has a resolution of the form
\eqn\free{ 0 \to V \to E_1 {\buildrel d_1 \over\longrightarrow} E_2
{\buildrel d_2 \over\longrightarrow} E_3 \to \cdots}
which is an exact sequence
and the $E_l$s are direct sums of line bundles.
The map
from $E_l$ to $E_{l+1}$ depends on $\phi$ and we will call it $d_l$.
Let $k_l$ be the rank of $E_l$.

At this point, the important point to
make is what goes wrong with the naive model when $d_1$
has a cokernel.  This is twofold: firstly, there are extra massless
$\wp$s on the boundary, each of which gives an energetically degenerate
harmonic-oscillator
spectrum of states, leading to the wrong
spectrum of CP factors.  Secondly, and perhaps more
importantly, a {\it generic} deformation of the matrix elements
of $d_1$ will destroy the brane.  This happens because
the matrix $d_1$ imposes overconstrained conditions on $\beta$, which
are inconsistent rather than redundant for a generic deformation.
So under a random deformation, the equations of motion for
the $\wp$s and $\beta$s will simply set them to zero at low energies,
meaning that no states will satisfy the boundary charge
projection -- there will be no CP factors.  This is connected
with the problem of massless $\wp$'s in that the superpartners
of the $\wp$'s are the Goldstone fermions for
the spontaneous breaking of supersymmetry.

Now for a model with the correct behavior.
The sections of the line bundles which we introduce will be alternately
Fermi multiplets, $\beta^{a}_{(l)}$, and chiral multiplets, $\wp_{a(l+1)}$.
They are
Fermi multiplets at the first step, as in the monad case, because it is
states of massless
fermions that play the role of the
CP factors (or the left-moving current algebra in a (0,2) theory).
The basic idea is that the fermions at the third step
pair up with the ``extra'' massless fermion partners
of the $\wp$'s at the second step (which
arise because $d_1$ has a cokernel).  If the sequence does not
terminate there, the partners of the bosons at the fourth step
pair up with the extra fermions at the third step, and this
process of lifting kernels and images continues until the
sequence terminates.

In the general case the boundary symmetry acts on all chiral multiplets
with one phase and all Fermi multiplets with the opposite phase.

For ease of discussion, we give the details for a bundle
with a two-step resolution:
\eqn\twostep{ 0 \to V \to E_1 {\buildrel d_1 \over\longrightarrow} E_2
{\buildrel d_2 \over\longrightarrow} E_3 \to 0 .}
We will also assume for simplicity that the line bundles making up $E_i$
at each step are the same, \ie,
$$
E\ll i \equiv \CO(n_i)^{\oplus k_i}.
$$
This assumption is in no way essential to the construction.

Then the degrees $\Delta_{1,2}$ of the polynomials defining the maps
$d_{1,2}$ are $n_2 - n_1$ and $n_3 - n_2$, respectively.
Here is the field
content:

$\bullet$ Bulk fields: the usual (2,2) multiplets ($\phi, P\ll {bulk}, \s$)
where $P\ll {bulk}$ is the \it bulk \rm $P$-multiplet.

$\bullet$ A set of boundary fermi multiplets $\b\uu i  $,
a smoothly varying subspace of which will define the desired bundle.
They satisfy the usual chiral constraint $\{ Q\dag, \b \} = 0$ and
their gauge charge is $n\equiv n_1$.

$\bullet$ A set of boundary chiral multiplets $\wp\ll {i }$
with gauge charges $ -n_2 = - (n+\Delta\ll 1) $.
Instead of the usual chiral constraint
$[Q\dag ,  \wp] = 0$ these will obey a deformed chiral constraint, which
will fix the $Q\dag$ variation of $\wp$ in terms of the other fields of
the problem.

$\bullet$
A set of boundary multiplets $\bprd$ which boundary-symmetry charge $+1$
and
gauge charge $n\ll 3 \equiv n + \Delta\ll 1 + \Delta\ll 2$.  They will
turn
out to be forced to satisfy the \it opposite \rm of the chiral constraint:
$\{Q, \b\pr\} = 0$.

Now for the interactions.
The half-superspace integral will be:
$$
\int d\th ~W\equiv \int d\th ~\wp d\ll 1 \b.
$$
We can add the obvious gauge-invariant kinetic term for
the $\b, \b\pr $ multiplets:
$$
\int d\sqd \th \cdot \b\dag\b + \b^{\prime\dagger}\b\pr
$$
The key to obtaining the correct physics is the multiplet structure of
$\wp$:

\bigskip
\it
Since $d\ll 1$ has a cokernel, the condition for the superpotential $W$ to
be annihilated by $Q\dag$ does not require $Q\dag$ to annihilate $\wp$.
Indeed,
$[Q\dag, \wp] = (anything)\cdot d\ll 2$ will suffice.
\rm
\bigskip

\noindent
There is an obvious choice: $[Q\dag, \wp] = i \beta^{\prime\dagger} d\ll
2$.
Since for a two-step sequence $d\ll 2$ is onto, the nilpotence of $Q\dag$
forces $\{ Q\dag, \beta^{\prime\dagger} \} = 0$ -- so $\beta\pr$ satisfies
the opposite of the usual chiral constraint.

This constraint completely determines the supersymmetry transformations of the new
multiplet.  This is one of the two key ingredients in this new type of
model.  The second is the observation that for a boson living on the
boundary, it is consistent to add to the Lagrangian
a magnetic field coupling for the complex target space fiber coordinates:
\eqn\susskind{
\int d\sqd \theta ~ \wp \wp^\dagger.
}
In fact it is necessary to add this term to obtain
the mass terms and Yukawa couplings that give the desired physics in
the infrared.

Rather than listing all the terms in the action and the supersymmetry transformations
here, we leave that to the Appendix.
In this section
we will discuss only the terms which will be important for causing the model
to flow to the correct brane configuration at low energies, in the 'large
radius' phase of the worldsheet theory.

To clarify the content of the low energy theory we begin by
integrating out the auxiliary fields $b$ in the fermi multiplets and also
the superpartners $\xi$ of $\wp$, which become auxiliary
at low energies in the presence of the
large magnetic field.  With auxiliary fields
eliminated, the key terms in
the Lagrangian are then:

$\bullet$ Derivative terms for the physical bosons and fermions of the system:
$$
i \cdot \lsq \b\dag (\nabla\ll 0 \b) + (\nabla\ll 0 \Pt) \Pt\dag
+ \bpr (\nabla\ll 0 \bprd ) - h.c. \rsq
$$

For the fermions these are just standard kinetic terms; for the bosons the
standard kinetic term is irrelevant and their dynamics at this scale is
dominated by lowest Landau level physics; the number of physical
degrees of freedom is effectively reduced by a factor of
two and the complex bosonic fiber becomes a product of noncommutative
$\IC\uu 1$'s.
Now the boson $\wp$ really does have similar kinematics to the
fermions $\b$ and $\bprd$, except with opposite statistics, i.e.,
$[\wp\ll i, \wp\dag\ll j] = \d\ll{ij}$.

$\bullet$ Mass terms for the fermions:

\eqn\fermmass{
\b\dag (d\ll 1\dag d\ll 1)\b + \bpr (d\ll 2 d\dag\ll 2 ) \bprd
}

$\bullet$ Mass terms for the bosons:

\eqn\bosmass{
\Pt (d\ll 1 d\dag\ll 1 + d\dag\ll 2 d\ll 2)
\Pt\dag
}

Note that because of the large magnetic field,
the physical mass of the bosons is proportional
to the coefficient of $|\wp|^2$ rather than to the
square root of that coefficient.

In the next subsection we explain why these potentials and Yukawa couplings
are preicisely what we need to yield the correct sigma model at low energies.

Interestingly, the supersymmetry
transformations take a very nice form with
$\xit$
and the other auxiliary fields integrated out:
\eqn\finalsusy{
\eqalign{
\{ Q, \b\} = - i d\dag\ll 1 \Pt\dag \hsk \{Q,\b\dag\} = 0 \cr
\{ Q\dag, \b\} = 0 \hsk \{ Q\dag, \b\dag\} = + i \Pt d\ll 1 \cr
[Q, \Pt\dag ] = + i d\dag\ll 2 \bprd \hsk [Q, \Pt] = - i \b\dag d\dag\ll 1
\cr
[Q\dag, \Pt\dag] = + i d\ll 1 \b \hsk [Q\dag, \Pt] = - i \bpr d\ll 2 \cr
\{Q,\bprd \} = 0  \hsk \{ Q, \bpr \} = - i \wp d\dag\ll 2 \cr
\{ Q\dag, \bprd\} = + i d_2 \Pt^\dagger  \hsk \{ Q\dag, \bpr\} = 0
}}
We note again that these transformations are consistent
with the supersymmetry algebra (in particular, $Q^2 = 0$)
because $d_2 \circ d_1 = 0$.  The relation $\{Q, Q^\dagger\} = 2 H$
holds on-shell, that is when the fields satisfy their (first-order) equations
of motion.  Although the limit in which the two-derivative kinetic term for
the bosons vanishes simplifies the algebra greatly, the model is consistent
without using this approximation.

The generalization to sequences of an arbitrary number
of steps should be clear.  The $Q\dag$-variation of a field
associated with a given node encodes the previous map, while
the $Q$-variation encodes the next map.  So, for example,
to make a 3-step sequence, we would add some $\wp\pr$ fields satisfying
the undeformed chiral constraint $[Q\dag, \wp\pr] = 0$ and
deform the constraint on $\b\pr$ to $\{Q, \b\pr\} = d\ll 3\dag
\wp\uu{\prime
\dagger}$.  The model works for a sequence of arbitrary
length; one never needs
to add another superpotential term, only deform the chiral constraint
by hand at each step according to the data of the sequence and add the
appropriate full-superspace terms.  In the appendix
we write down the model for a sequence of arbitrary length.

\subsec{Large-radius analysis}

The point of this section is to prove that we get the right
low-energy behavior for the two-step model
from the interactions discussed in the previous
subsection.
Make the bulk FI coefficient $r$ large and positive,
so that we are in the large-radius CY phase of the bulk theory.

Let us determine the supersymmetric vacuum of the theory
by setting to zero the supersymmetry variations in equation \finalsusy :
\eqn\vaceqn{ \eqalign{
0 &= d_1  \beta \cr
0 &= \wp  d_1 \cr
0 &= \wp  d_2^\dagger \cr
0 &= d_2^\dagger  \bprd
}}
The first equation tells us that the massless $\beta$s live in the
kernel of $d_1$.  Since $d_2$ is surjective, the last equation tells
us that $\bprd$ must vanish.  The two middle equations tell us that
$\wp$ is closed and co-closed which, since the sequence is exact
at $E_2$, tells us that $\wp$ must vanish as well.  This is
the desired physics.

We would arrive at the same conclusion by a direct examination of the
Lagrangian, without making use of supersymmetry.  The statement that the
sequence we examine has no cohomology at the middle node is the statement
that there is no nonzero fiber annihilated both by $d_2$ and by $d_1^\dagger$.
As a result, $d_2^\dagger d_2 + d_1 d_1^\dagger$ is an invertible matrix,
so all components of $\wp$ are set
to zero at low energies by their equations of motion.
Similar reasoning shows that all $\beta^\prime$ are set to zero by their
mass terms at low energies, and that the surviving subset of the $\beta$'s
is the kernel of $d_1$, just as we wanted.

Note that this analysis makes it clear that
there is a direct relation between cohomology of the sequence and massless
worldsheet fields.
In particular, if under a deformation the defining polynomials the
sequence fails to
be exact at some point, $\phi_\star$, in the CY, closed and co-closed will
no
longer imply zero and we will find massless $\bprd$s and
$\wp$'s.

In summary, as in the simpler case, given a non-degeneracy condition
for the sequence
(analogous to $G = d_1 = 0$ has no solutions in the
monad case \dk)
all of the $\wp$'s vanish in
vacuum, and the mass matrix for the fermions imposes the sequence.

\subsec{Examples}

The first example we study is a very trivial one,
namely a multi-step resolution for a twobrane on a two-torus with
a trivial line bundle.  Make the $T^2$ as a cubic hypersurface
in $\IP^2$.  Consider the Koszul complex over the ambient
$\IP^2$,
$$
0 \to V~ {\buildrel i \over\hookrightarrow} ~\CO(1)^{\oplus 3}~
{\buildrel d_1 \over\longrightarrow}
~\CO(2)^{\oplus 3}~
{\buildrel d_2 \over\longrightarrow} ~\CO(3) \to 0 $$
where the maps are
$$ d_1 = \left (
\matrix{ 0 & \phi_2 & -\phi_1 \cr -\phi_2 & 0 & \phi_0 \cr \phi_1 & -\phi_0 & 0 } \right) ~~~
{\rm and} ~~~
d_2 = \left( \phi_0, \phi_1, \phi_2 \right) ;$$
the inclusion is induced by
$$ i = \left( \matrix{ \phi_0 \cr \phi_1 \cr \phi_2 } \right).$$
One can see by computing its Chern classes
that the line bundle $V$ defined
by this sequence is in fact trivial\foot{This statement is false.
The example of the CY 1-fold is
special in that the target space is not simply connected; however
the existence of a global section (demonstrated below) means
that the bundle is in fact trivial.}.

To model a string coupling to this brane, we add on its boundary
three fermi fields $\beta$ of charge $1$, three chiral fields
$\wp$ of charge $-2$, and another fermi field $\bprd$ of charge $3$.
We add the superpotential
$$ W = \wp d_1 \beta = \epsilon_{ijk} \beta^i \wp^j \phi^k $$
and implement the chiral constraints and charge
projections discussed above.

Obviously this brane could also be constructed by simply adding a neutral
fermion on the end of the string.
To see the relation to the above model,
define an effective neutral fermion, $\gamma$, by
$$ \beta^i = \phi^i \gamma + {\rm massive}.$$
Setting to zero $\wp, \bprd$ and the massive components of $\beta$ then
solves the vacuum equations identically.  So
the multistep model does flow in the IR to the
theory of a single neutral fermion.

For a less trivial example, we study
the pullback to the CY of the tensor square
$V \equiv {\rm T}^* \IP^n \otimes {\rm T}^* \IP^n$
of the cotangent bundle of
$\IP^n$.  This is defined by the sequence
$$
0 \to V~ {\buildrel i \over\hookrightarrow} ~\CO(-2)^{\oplus (n+1)^2}~
{\buildrel d_1 \over\longrightarrow}
~\CO(-1)^{\oplus 2n+2}~
{\buildrel d_2 \over\longrightarrow} ~\CO(0) \to 0
$$
where the maps are
$$ d_1 : \beta_{ij} \mapsto \left(\phi^j \beta_{ij} , \phi^i \beta_{ij} \right)
~~~ {\rm and} ~~~
d_2: \left(\wp, \hat{\wp} \right) \mapsto \phi^i (\wp_i - \hat{\wp_i}) .$$

To build this bundle, we add $(n+1)^2$ fermi multiplets $\beta_{ij}$ of
charge $-2$, $2n+2$ chiral multiplets $\wp_i$ and  $\hat{\wp_i}$
of charge $-1$, and one neutral fermi multiplet $\bprd$.
The superpotential is
$$ W = \wp_i \phi^j \beta_{ij} + \hat{\wp}_j \phi^i \beta_{ij} $$
and the nontrivial deformed constraints are
$$
\eqalign{
[Q, \wp_i] = -i \beta_{ij}\dag \phi^{j \dagger} ~~~
&[Q, \hat{\wp}_j ] = - i \beta_{ij}\dag \phi^{i \dagger} \cr
[Q\dag, \wp_i] = -i \bpr \phi^i ~~~
&[Q\dag, \hat{\wp}_i] = i \bpr \phi^i \cr
\{ Q, \bpr \} = &- i (\wp - \hat{\wp})\dag \phi^{i \dagger} .}
$$

We have given two examples of smooth bundles
with constant fiber dimension using resolutions
of finite length.  It would be nice to understand
more about the physics of the sheaves defined by
more general choices of
ranks and charges in a multistep sequence \free.

\newsec{Bundles which do not extend to the ambient space}

In fact we lose some generality by considering only
pullbacks of bundles on the ambient toric variety.
Bundles which extend generate in general a sublattice of finite index in
the lattice of all $K$-theory classes of bundles on the target variety.

In order to see how to generate linear models for more general bundles, let
us consider the case of a hypersurface defined by $G(\phi) = 0$.  The bulk
then has a single $P$-field and a superpotential $W\ll{\rm bulk} \equiv
P G(\phi)$.  Non-extending bundles can be realized as the
cohomology of a sequence {\it over the coordinate ring of the variety},
i.e.
the cohomology of a set of maps $d\ll n$ such that $d\ll{n+1} d\ll n
= M\ll{n+1 | n} G(\phi)$ for some matrix $M\ll{n+1 | n} $ of holomorphic
polynomials.

The key fact in the following construction is that
$$
d\ll {n+2} M\ll{n+1 | n} = M\ll{n+2 | n+1} d\ll n
$$

\it Proof: $G(\phi) (d\ll {n+2} M\ll{n+1 | n} - M\ll{n+2 | n+1} d\ll n)
= (d\ll {n+2} d\ll {n+1}) d\ll n - d\ll {n+2} (d\ll {n+1} d\ll n) = 0 $ by
associativity.
Since $G$ is nonvanishing and polynomial rings contain no zero divisors,
that means the second factor $d\ll {n+2} M\ll{n+1 | n}-M\ll{n+2 | n+1} d\ll n$
must vanish and the statement follows. \rm

We now set the bulk $\eta$ multiplet on shell; in particular this means
$ {F^\dagger}^P \equiv
\{ Q\dag, {\eta^\dagger}^P\} = G(\phi)$.  For a two-step
resolution we take the boundary superpotential to be
$$
W = (\wp d\ll 1  - {\eta^\dagger}^P \bpr M\ll{2|1} ) \b
$$
and the (deformed) chiral constraints to be
$$
\{Q\dag, \b\} = 0 ~~~~ [Q\dag, \wp ] = \bpr d\ll 2
~~~~
\{Q\dag, \b\pr\} = i d_2 \wp \dag.
$$
Clearly the supersymmetry algebra closes and the superpotential is annihilated by
$Q\dag$.

The extension to a multistep sequence over a hypersurface is straightforward.
For three steps, for instance, the superpotential is the same, and the
supersymmetry transformations are
$$
\eqalign{
\{Q\dag, \b\} &= 0 \cr
[Q\dag, \wp ] &= \bpr d\ll 2 - {\eta^\dagger}^P \wp\pr M\ll{3|2} \cr
\{Q\dag, \bprd\} &= i d_2 \wp\dag ~~~~~  \{Q\dag, \bpr\} = \wp\pr d\ll 3  \cr
[Q\dag, \wp\pr] &= 0 }
$$

For an arbitrary number of steps, take the same superpotential and
use the constraints
$$
\eqalign{
\{Q\dag, \beta \} &= 0 \cr
[Q\dag, \wp ] &= -i \bpr d\ll 2 - {\eta^\dagger}^P \wp\pr M\ll{3|2} \cr
\{Q\dag, \bpr\} &= i \wp\pr d\ll 3 - {\eta^\dagger}^P \b^{\prime\prime\dagger}
 M\ll{4|3}  \cr
&\cdots }
$$

\newsec{Discussion}
\subsec{Relation to heterotic models}

\hbox{\vtop{\hsize=2.625in
\def\tablerule{\omit&
\multispan{8}{\tabskip=0pt\hrulefill}&\cr}
\def\tablepad{\omit&
height3pt&&&&&&&&\cr}
$$\vbox{\offinterlineskip\tabskip=0pt\halign{
\hskip-.25in
\strut$#$\ \ &
\vrule#&\ \ \hfil $#$ \hfil\ \ &\vrule #&\ \
\hfil $#$ \hfil\ \ &\vrule #&\ \
\hfil $#$ \hfil\ \ &\vrule #&\ \
\hfil $#$ \hfil\ \ &\vrule #\cr
&\omit&\hbox{Field}&\omit&q_G&\omit&q_L&\omit&q_R&\omit\cr
\tablerule\tablepad
&&\Phi_i&& w_i && 0 && {w_i \over m}&\cr
\tablepad\tablerule\tablepad
&&\Gamma&& - d&&0 && 1-{d \over m}&\cr
\tablepad\tablerule\tablepad
&&\beta&&n_1&& 1 && {n_1 \over m}&\cr
\tablepad\tablerule\tablepad
&&\wp&&-n_2&&-1 && 0&\cr
\tablepad\tablerule\tablepad
&&\beta^{\prime}&&n_3&&1&& 1 - {d_2 \over m}& \cr
\tablepad\tablerule
\noalign{\bigskip}
\noalign{\narrower\noindent{\bf Table 1:}
The gauge charges and ``left-moving $U(1)$'' charges of the fields.}
 }}$$}
\vtop{\advance\hsize by -2.625in
One might hope that one could use similar technology
to make linear models for heterotic strings coupling
to these bundles.  In fact, our spectrum of fields
was motivated by the anomaly coefficients and
cancellation conditions that one would have in the
heterotic case.  In the table at left we write the
gauge charges, boundary symmetry charges, and
R-symmetry charges of our fields for a generic
two-step model.
}}
\noindent
Now imagine that
they are instead representations of (0,2) supersymmetry,
as in \dk.
We call the boundary symmetry charge $q_L$ because
in its heterotic incarnation, it is the charge under
the left-moving
$U(1)$ which becomes part of the spacetime gauge group
(and a $Z_2$ subgroup of which provides one of the
GSO projections).
Using these charges we would
calculate the anomaly in the left-moving $U(1)$ to be
$$ \CA(L, G) \propto \sum_{fields} q_G q_L = k_1 n_1 - k_2 n_2 + k_3
n_3.$$
This is just $c_1(V)$
\foot{The chern classes of the bundle $V$ are determined by the sequence
\free\ to be
$$ c(V) = \prod_{l = 0} c(E_l)^{(-1)^l} $$
and in particular
$ c_1(V) = \sum_l (-1)^l \sum_{a_l} n_{a_l} J$,
and
$ c_2(V) = \sum_l \sum_{a_l \neq a_l'} n_{a_l} n_{a_l'} J^2 $
where $J$ is the $(1,1)$ form on the CY.}.  Note that
consistency only requires this to vanish mod 2, since only a
$Z_2$ subgroup of $U(1)_L$ is gauged.
The gauge anomaly would be
$$ \CA(G, G) \propto \sum_{fields} (-1)^{fermi} q_G q_G \propto
c_2(V) - c_2(X) .$$
The anomaly in the $U(1)_R$ symmetry would be (modulo the
gauge anomaly)
$$ \CA(R, G) \propto \sum_{fields} q_G q_R = \sum_i w_i - d,$$ the first
chern class of the hypersurface.

As in \refs{\dk, \trieste}
we can calculate the left-moving central charge of our model
as an anomaly-matching coefficient in the massive theory.  Specifically,
the left-moving central charge is the would-be (if it were gauged)
quadratic anomaly of the left-moving
$U(1)$ symmetry.  This is $r = k_0 - k_1 + k_2$, the rank of
the bundle, as expected.

However, while the spectrum seems to be correct, the interactions
we would add to give the right vacuum structure
do not respect two-dimensional Lorentz invariance.
An equivalent way to phrase the obstruction is that the
desired interactions break the $U(1)$ R-symmetry
which is thought to flow to the R-symmetry of the
right-moving $\CN=2$ superconformal algebra.
It is possible that a new R-symmetry appears in
the IR, but without a UV candidate for the IR R-symmetry
one loses some confidence that
one has identified the correct LSM.
This is puzzling, but
we still hope to find a LSM for the multi-step resolution in the
heterotic case.
Given that
nonperturbative conformal invariance of (0,2) models
has only been proven using the linear models \evaed, it would be
fascinating if
no heterotic LSM could be found for multi-step bundles.

We also hope to find an argument that motivates the field content directly in
the open string case, such as a direct relation between the RR charge
of the D-brane configuration and the spectrum of worldsheet fields.

\subsec{Other phases and singularities}

One of the great successes of the linear sigma model approach
to CY physics is an automatic description of the
stringy physics of small-volume phases of the theory.
We are still working out the behaviour of the
monad theories when the FI parameter is large and negative \hklm,
and
we leave the application of that analysis to these more intricate models
for future work.

But there is a possibility of new physics from our multistep bundles.
When the moduli of the bundle are deformed in such a way that the
maps beyond the first step degenerate, cohomology will appear
at higher nodes.
It would be interesting to see if the resulting
massless worldsheet fields have any special signatures in the spacetime physics.

We have been assuming thoughout this paper that the sequence
\eqn\noV{ 0 \to E_1 {\buildrel d_1 \over\longrightarrow} E_2
{\buildrel d_2 \over\longrightarrow} E_3 \to \cdots}
was chosen so that its cohomology was a bundle.
In fact, the cohomology of such a sequence will generically
be a more general coherent sheaf
(espoused by Harvey and Moore \bpsalg\ as the proper mathematical
characterization of wrapped D-branes).  Work is in progress to harness this fact to study
D-brane configurations without space-filling branes \hklm.

\subsec{Other classes of models}

In addition to the heterotic models discussed above, we are hoping to extend the construction to:

\item{1.} (2,2) linear models for varieties which are not complete intersections.
The generic CY manifold is such a beast.  If a variety is not a complete intersection,
it means that the number of defining equations is bigger than its codimension.
As a result, there are relations among these equations, and in general relations
among these relations...  There is again a sequence of maps resolving the
ideal of the variety.  We have made some progress towards such (2,2) models
using extra gauge symmetries.
\item{2.} Open strings in the presence of branes wrapping submanifolds which
are not complete intersections.

\bigskip
\noindent
If we can accomplish item 1 above, our technology will
be very useful for the program of \recentmrd.  In particular,
the 4d $\CN=1$ field theory which is proposed to describe
the space of branes with fixed charge involves a
superpotential with relations among the vacuum equations,
and relations among these relations $\dots$

\appendix{A}{{Full supersymmetry transformations and lagrangian}}

In this appendix we give a more detailed
description of the supersymmetry structure of our models,
and present the model for a sequence with an
arbitrary number of nodes.

First we present the gauge-covariantized
supersymmetry transformations for the two-step case
and show that we get all of the
terms we need in the action
to get the desired massless degrees of freedom.
(We do
not write down
terms in the bulk action
which are given in \refs{\phases}.)  We suppress indices on the fields.

The transformations of the bulk multiplets under the reduced superalgebra
are:
$$
\eqalign{
[Q,\phi ] &= - i \Theta  \hsk [Q\dag, \phi  ] = 0 \cr
\{Q, \Theta\} &= 0 \hsk \{Q\dag, \Theta \} = 2 \nabla\ll 0 \phi }
$$
and
$$
\eqalign{
\{Q,\eta \}  &= F  \hsk \{Q \dag, \eta  \} =2 \tilde\nabla_1 \phi\dag \cr
[Q, F ] &= 0 \hsk
[Q \dag, F] = -2 i \nabla_0 \eta + 2 i \tilde \nabla_1 \Theta +
2 i q \phi \left( \lambda_+\dag + \lambda_- \dag \right).
}
$$
where $\tilde \nabla_1 \equiv {i \over 2 \sqrt 2 } \{Q , Q_+\dag + Q_- \dag\}$,
the anticommutator of $Q$ with the broken supercharge.

We will introduce the notation $\xi$ for the fermionic superpartner of
$\wp$.  So:
$$
\{Q, \bpr\} = b\pr \hsk \{Q, \bprd \} = 0
$$
$$
\{ Q\dag, \bpr\} = 0 \hsk \{Q\dag, \bprd \} = b\uu
{\prime\dagger}
$$
$$
[Q, b\pr] = 0 \hsk [Q, b\uu{\prime\dagger} ] = - 2 i \nabla\ll 0 \b\pr
$$
$$
[Q\dag, b\pr] = - 2 i \nabla\ll 0 \bpr \hsk [Q\dag, b\uu{\prime\dagger} ]
=
0
$$
$$
[ Q, \Pt ] = - i \xit \hsk [Q, \Pt\dag] =
+ i d\ll 2\dag \bprd
$$
$$
[Q\dag, \Pt] = + i \bpr d\ll 2 \hsk
[Q\dag, \Pt\dag] = - i \xit\dag
$$
$$
\{Q, \xit \} = 0 \hsk \{Q, \xit\dag\} =
2 \nabla\ll 0 \Pt\dag + d\ll 2\dag b\uu{\prime\dagger} -
i (d\ll{2,a}\dag \theta\uu{a\dagger} ) \bprd
$$
$$
\{Q\dag, \xit \} = 2 \nabla\ll 0 \Pt
 + b\pr d\ll 2 + i \bpr (d\ll{2,a}
\theta\uu a) \hsk \{Q\dag, \xit \dag\} = 0
$$

So the $d\sqd\th$ integral of $\bpr \bprd$ is
$$
b\pr b\uu{\prime\dagger}  - i (\nabla\ll 0 \bpr) \bprd
+ i \bpr (\nabla\ll 0 \bprd )
$$
and the $d\sqd\th$ integral of $\Pt \Pt\dag$ contains five types of term:

$\bullet$ a 'target space magnetic field' term
\eqn\magfield{
+ i (\nabla\ll 0 \Pt) \Pt\dag - i \Pt(\nabla\ll 0 \Pt\dag)
}

$\bullet$ quadratic terms for the gauge-invariant $\xi$ fermions:
\eqn\fermaux{
- \xit\xit\dag
}

$\bullet$ mass terms for the $\bprd$ fermions:
\eqn\bprdmass{
+ \bpr d\ll 2 d\ll 2\dag \bprd
}

$\bullet$ some F-terms giving rise to a bosonic potential for $\Pt$:
\eqn\Fterm{
i b\pr d\ll 2 \Pt\dag - i \Pt d\dag\ll 2 b\uu{\prime\dagger}
}

$\bullet$ and some cross terms between $\bprd$ and $\Pt$ which have no obvious
role:
\eqn\useless{
- \bpr (d\ll {2,a} \theta\uu a) \Pt\dag - \Pt (d\dag\ll{2,a}
\theta\uu{a\dagger})
\bprd
}

$\bullet$ We also have the usual kinetic terms for the $\b$-fermions:
\eqn\kinterms{
\int d\sqd \th \b\dag\b = + b\dag b  - i (\nabla\ll 0 \b\dag)
\b
+ i \b\dag (\nabla\ll 0 \b )
}

\noindent
Also, the superpotential contributes the following component terms:
\eqn\supterms{
\eqalign{
\int d\th W &\equiv i \int d\th \wp d\ll 1\b = i\int d\th \Pt d\ll 1\b \cr
&= \xit d\ll 1 \b  + \b\dag d\dag\ll 1 \xit\dag \cr
&+ \Pt (d\ll {1,a} \theta\uu a ) \b + \b\dag (d\dag\ll
{1,a}\theta\uu{a\dagger})
\Pt\dag \cr
&+  i \Pt d\ll 1 b - i b\dag d\dag\ll 1 \Pt\dag.
}}

\bigskip
\noindent
{\it The model with many nodes}

For an arbitrary number of steps,
the action is
$$
\int d^2 \theta \sum_n \left( \beta\dag_{(n)} \beta_{(n)} +
\wp_{(n+1)} \wp_{(n+1)}\dag \right) + \int d\theta ~W + {\rm h.c.}
$$
and the on-shell
supersymmetry transformations are
$$
\eqalign{
\{Q\dag, ~\beta_{(1)} \} = 0 ~~&~~~
\{Q, \beta_{(1)} \} = - i d_1\dag \cdot \wp_{(2)}\dag \cr
& \cdots \cr
\{Q\dag, ~\beta_{(n)} \} = i d_{n-1} \cdot \wp_{(n-1)}\dag ~~&~~~
\{Q, ~\beta_{(n)} \} = - i d_{n}\dag \cdot \wp_{(n+1)}\dag \cr
[Q\dag, ~\wp\dag_{(n+1)} ] = i d_{n} \cdot \beta_{(n)} ~~&~~~
[Q, ~\wp\dag_{(n+1)} ] = i d_{n+1}\dag \cdot \beta_{(n+2)} \cr
& \cdots
}$$

\appendix{B}{{Formulation with shift symmetries}}

In the body of this paper we have given a presentation
of our models which clearly exhibits the spectrum of fields and their
interactions.  This presentation has the drawback that the
supersymmetry transformations of the fields are complicated and
the closure of the supersymmetry algebra is not manifest.  In
this appendix we present a formulation of our models in which
the supersymmetry transformations of all fields are simple but
which involves a number of nonlinearly-realized gauge symmetries.
The formulation of the models given in the body of the paper
results from fixing these gauge symmetries.
This appendix clarifies the relation between the
modified chiral constraints and
the
``fermionic gauge symmetries''
of \refs{\dk, \trieste}, and is how our models were initially
constructed.

We again present the construction for a two-step sequence, \twostep.
Introduce a fermi multiplet $\beta_1$ of bulk gauge charge
$n_1$, a chiral multiplet
$\tilde \wp_2$ of bulk gauge charge $-n_2$, and an unconstrained
multiplet $V_3$ of gauge charge $n_3$ whose lowest component is a boson.
The usual superpotential
$$ W = \tilde \wp_2 d_1 \beta $$
respects the following
shift symmetry, which we gauge:
$$
\eqalign{
\tilde \wp_1 &\mapsto \tilde \wp_1 + \Omega_2 d_2 \cr
V_3 &\mapsto V_3 + \Omega_2
}$$
The gauge parameter $\Omega_2$ is a chiral multiplet,
$Q\dag \Omega_2 = 0$.  Then the shift-symmetry invariant field
$\wp_2 \equiv \tilde \wp_2 - V_3 d_2$ satisifes the constraint
$ Q\dag \wp_2 = \beta_3 \dag d_2 $ with $\beta_3\dag  \equiv - Q\dag V_3$.
$\beta_3 \dag $ is gauge invariant and satisfies $Q \dag \beta_3 \dag = 0$.

In order to extend the sequence by another step, enlarge the gauge
symmetry.  Relax the condition $Q^\dagger \Omega_2 = 0$
on the gauge parameters and
demand merely
that $Q^\dagger \Omega_2 = \Omega_3 d_3$ for some $\Omega_3$
with $Q^\dagger \Omega_3 = 0$.
Note again that the fact that the maps $d_l$ form a complex
is crucial for closure of the superalgebra.
Under this modified shift symmetry, the quantity $\wp_2$ is still gauge
invariant, but $\tilde{\beta_3}\dag   \equiv - Q^\dagger V_3$ is
not.
Introduce a new gauge (\ie\ unconstrained) multiplet
$V_4$ (whose lowest component
is a fermion)
and assign it a transformation
$\Sigma_4 \mapsto \Sigma_4 +
\Omega_3$.  Then $\beta_3^{\dagger} \equiv
\tilde{ \beta_3 }^{\dagger} - \Sigma_4 d_3$
is gauge invariant, and satisfies the deformed chiral
constraint $Q^\dagger \beta_3^{\dagger} = \wp_4 d_3$
where
$\wp_4 \equiv - Q^\dagger \Sigma_4$ is gauge invariant and
annihilated
by $Q^\dagger$.  The generalization of this formulation
to an arbitrary number of steps should be clear.

\bigskip

\centerline{\bf{Acknowledgements}}
We thank Paul Aspinwall, Sarah Dean, Jacques Distler,
Shamit Kachru, Sheldon Katz,
Albion Lawrence, Dave Morrison and Eva Silverstein for discussions.
The work of Simeon Hellerman is supported by the DOE
under contract DE-AC03-76SF00098 and
by a DOE OJI grant.

\listrefs
\end